# Gene expression and pathway bioinformatics analysis detect a potential predictive value of MAP3K8 in thyroid cancer progression

Valentina Di Salvatore
Dept. of Biomedical and Biotechnological Sciences
University of Catania
Catania, Italy
valentina.disalvatore@unict.it

Fiorenza Gianì
Dept. of Department of Clinical and Molecular
BioMedicine, Garibaldi-Nesima Medical Center
Catania, Italy
fiorenza.giani@gmail.com

Giulia Russo
Dept. Of Drug Sciences
University of Catania,
Catania, Italy
giulia.russo@unict.it

Marzio Pennisi
Dept. of Mathematics and Computer Science
University of Catania,
Catania, Italy
mpennisi@dmi.unict.it

Pasqualino Malandrino
Dept. of Clinical and Experimental Medicine
University of Catania
Catania, Italy
linomalandrino@gmail.com

Francesco Frasca
Dept. of Department of Clinical and Molecular
BioMedicine, Garibaldi-Nesima Medical Center
Catania, Italy
f.frasca@unict.it

Francesco Pappalardo
Dept. Of Drug Sciences
University of Catania,
Catania, Italy
francesco.pappalardo@unict.it

Corresponding author: Francesco Pappalardo

*Abstract*— Thyroid cancer is the commonest endocrine malignancy. Mutation in the BRAF serine/threonine kinase is the most frequent genetic alteration in thyroid cancer. Target therapy for advanced and poorly differentiated thyroid carcinomas include BRAF pathway inhibitors. Here, we evaluated the role of MAP3K8 expression as a potential driver of resistance to BRAF inhibition in thyroid cancer. By analyzing Gene Expression Omnibus data repository, across all thyroid cancer histotypes, we found that MAP3K8 is up-regulated in poorly differentiated thyroid carcinomas and its expression is related to a stem cell like phenotype and a poorer prognosis and survival. Taken together these data unravel a novel mechanism for thyroid cancer progression and chemo-resistance and confirm previous results obtained in cultured thyroid cancer stem cells

## I. INTRODUCTION

Thyroid cancer is the most common type of endocrine tumor showing an increasing incidence over the last three decades. Malignant carcinoma of the thyroid arises from two different cell types, follicular and parafollicular. Follicular cells are involved in the production of thyroid hormones and they may give rise to well-differentiated and anaplastic thyroid carcinomas. The parafollicular C cells are responsible for the calcitonin production and they may give rise to medullary thyroid cancer (MTC) [1].

BRAF participates in the production of a protein involved in the transmission of chemical signals from outside the cell to the nucleus. This protein is part of the RAS/MAPK signaling pathway, which controls several important cell functions. Specifically, the RAS/MAPK pathway regulates cell proliferation, differentiation, migration and the programmed death (apoptosis) [2].

Among the various Raf kinase isoforms, the B-type RAF V600E (BRAFV600E) mutation is the most commonly observed inducing excessive proliferation and differentiation of tumor cells at the initial tumor stage [3].

Mutations in the BRAF serine/threonine kinase represent the most common genetic cause in thyroid cancer, occurring in approximately 45% of papillary thyroid cancer (PTC) and in a lower proportion of poorly differentiated thyroid cancer (PDTC) and anaplastic thyroid cancer (ATC) [4].

Moreover, BRAF gene mutations may be considered as a predictive factor for lymph node metastasis, extrathyroid extension, advanced disease stages III and IV, and disease recurrence [5].

The importance of the MAPK pathway has been well established in the tumorigenesis of PTC. The MAPK pathway is driven by activating mutations, including BRAF and RAS mutations, RET/PTC, TRK and ALK rearrangements. MAPK-mediated thyroid tumorigenesis involves a wide range of secondary molecular alterations that synergize and amplify the oncogenic activity of this pathway, such as genome-wide hypermethylation and hypomethylation and altered expression of miRNAs. Upregulation of various oncogenic proteins can occur and drive cancer cell proliferation, growth, migration and survival, as well as tumor angiogenesis, invasion and metastasis [6].

In a previous work we found that thyroid cancer stem cells derived from 8505 cell line are resistant to the BRAF inhibitor vemurafenib, despite harboring BRAFV600E mutation. In these cancer stem cells the resistance to vemurafenib was mediated by a paradoxical over-activation of ERK and AKT pathways. By our computational modeling, we found a fundamental role of mitogen-activated protein kinase 8 (MAP3K8), a serine/threonine kinase expressed in thyroid CSCs, in mediating this drug resistance. Hence, in this paper, we investigate the MAP3K8 expression in vivo in a large series of human thyroid cancer samples and its relationship to tumor behavior.

## II. MATERIALS AND METHODS

In order to analyse the trend of MAP3K8 expression values across all types of thyroid cancers (ATC, PTC, PDTC), the following datasets have been selected and downloaded from Gene Expression Omnibus data repository (https://www.ncbi.nlm.nih.gov/geo/):

- GSE33630 (11 ATC, 49 PTC and 45 normal samples)

- GSE76039 (17 PDTC and 20 ATC samples)

- GSE58545 (27 PTC and 18 normal samples)

- GSE3678 (7 PTC and 7 normal samples).

*A. Survival analysis*

Bioconductor survival package [7] has been used to obtain Kaplan-Meier survival curves: the Kaplan-Meier estimate is a nonparametric maximum likelihood estimate (MLE) of the survival function, S(t). This estimate is a step function with jumps at observed event times, ti. Kaplan-Meier curves show the survival probability over time of specific subjects under specific conditions, as shown in Figure 1. Survival analysis was performed on data relating to a cohort of patients suffering from papillary thyroid cancer (PTC), obtained from the GDC data portal (https://portal.gdc.cancer.gov/), within The Cancer Genome Atlas (TCGA) project . PTC is the most common form of thyroid cancer and since it is less aggressive than ATC it is very treatable, especially if diagnosed early and in people under 50. In order to produce Kaplan-Meier curves, both clinical and expression values data were used.

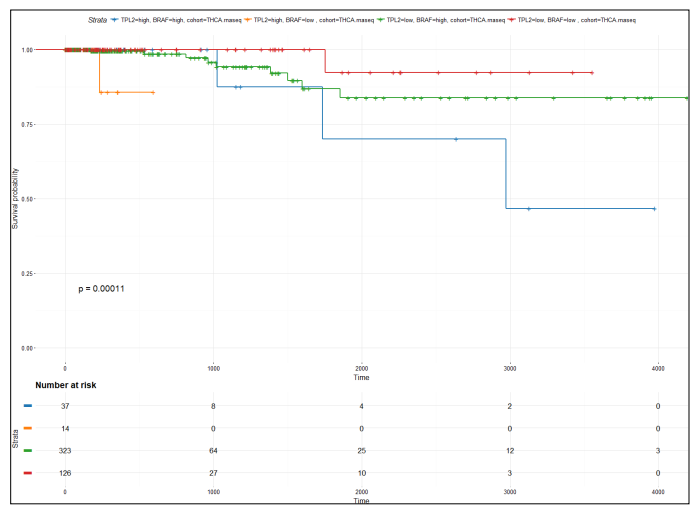

Figure 1. Kaplan-Meier curves show the survival probability of groups of patients with different BRAF and TPL2 expression levels.

*B. Gene expression analysis*

Raw signal intensities for each probe set contained in the CEL files were analyzed using different methods implemented in Bioconductor statistical programming language R (http://bioconductor.org). Bioconductor GEOquery package [8] were used for analyses of microarray data. GC-RMA algorithm was used for normalization process: GC-RMA algorithm used probe sequence information to estimate probe affinity to non-specific binding. Analysis was performed after normalization of the intensities from all of the chips into homogeneous distributions. The first step of analysis consisted in filtering out uninformative data such as control probesets and other internal controls and removing genes with low variance, that could have compromised the results of statistical tests for differential expression. Once datasets were filtered, data were sent to limma package for differential gene expression analysis [9].

*C. Pathway Analysis*

Bioconductor package SPIA [10] has been used to implement a Signaling Pathway Impact Analysis (SPIA) on our list of differentially expressed genes along with their log fold changes and KEGG signaling pathway topologies in order to identify most relevant pathways under our conditions of study. Among all the 136 pathways we obtained, we selected only those where our target gene MAP3K8 was involved in: *i)* MAPK signaling pathway; *ii)* Toll-like receptor signaling pathway; *iii)* T cell receptor signaling pathway. All the mentioned above pathways are respectively depicted in Figure 2, 3 and 4. Moreover, single nucleotide polymorphisms among various TLR genes have been identified, and their association with susceptibility/resistance to certain infections like pulmonary tuberculosis and other inflammatory diseases has been reported [11].

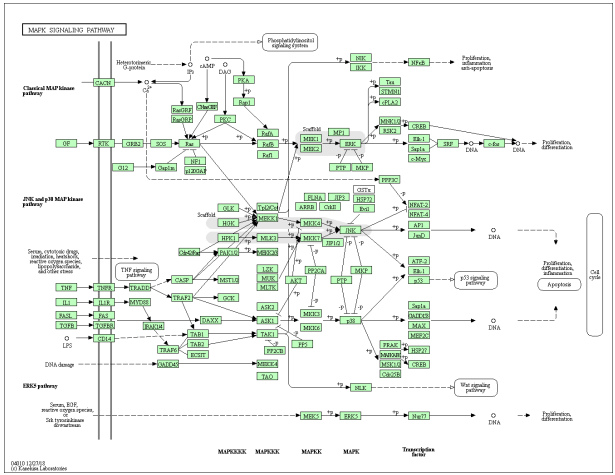

Figure 2. MAPK Signaling Pathway, MAP3K8 is in red (alias ID Tpl2/Cot).

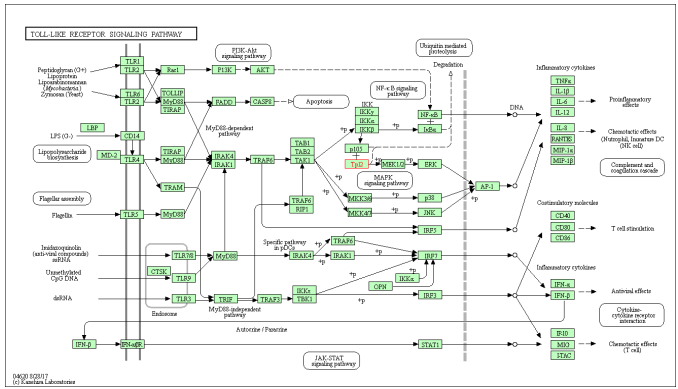

Figure 3. Toll-like receptor signaling pathway: Tpl2 target gene is highlighted in red.

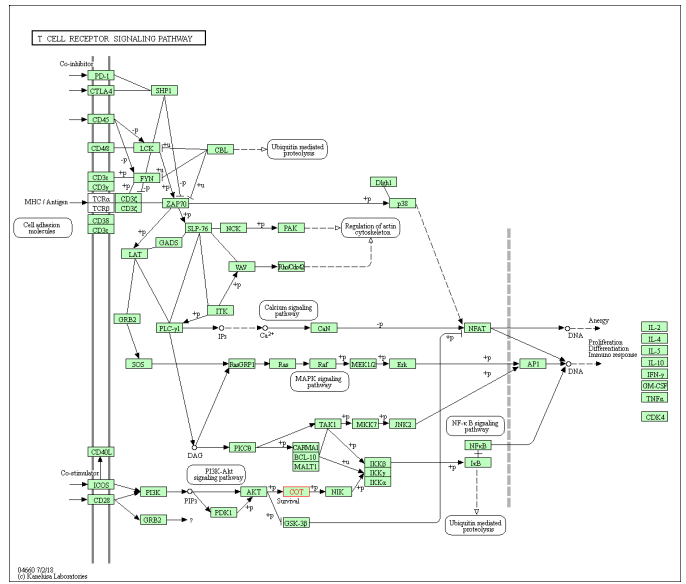

Figure 4. T cell receptor signaling pathway: COT is highlighted in red.

*D. Stemness index evaluation*

As cancer progression is accompanied by a gradual loss of a differentiated phenotype and the acquisition of stem-cell like phenotype, we assessed whether the stemness degree of a particular thyroid cancer type could be a predictive parameter of tumor outcome. Using a one-class logistic regression (OCLR) algorithm trained on stem cell, we derived mRNA expression-based stemness index (mRNAsi) on 37 samples of our selected dataset GSE76039 [12]. Stemness index values range from low (0) to high (1): after calculating mRNAsi for each sample of our reference dataset, we ranked our data depending on mRNAsi value thus obtaining two groups, one showing high index values and the other one showing low index values, using a 0.5 value as cut-off. Then, we repeated differentially expressed genes (DEG) analysis on each group separately looking for any correlation between high or low index values and MAP3K8 expression levels.

III. RESULTS

Differentially expressed genes (DEGs) analysis was performed on each dataset individually first, and then extended to all datasets simultaneously to compare expression values across all samples.

DEGs analysis provided a list with all genes displaying any changes in expression levels between normal and cancer samples. This list includes many genes whose involvement in many types of tumor formation is widely discussed in literature, as shown in Figure 5.

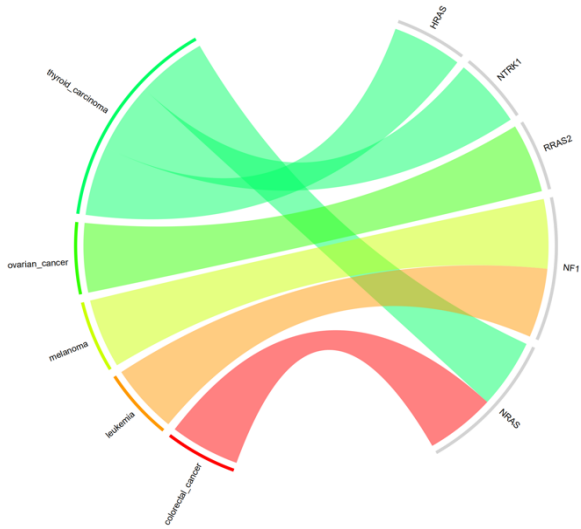

Figure 5. This graphic shows the correlation between different cancer types and some of the genes of our DEGs list.

Survival analysis was conducted in order to find out whether any alteration in differentially expressed genes expression values correlates with worse survival or earlier tumor recurrence. Survival curves, indeed, showed that death probability is higher in case of both BRAF mutation and MAP3K8 high expression levels, thus confirming MAP3K8 role as antagonist to BRAF inhibitor drugs. DEGs analysis conducted simultaneously on all Thyroid cancer types across all samples showed a MAP3K8 over-expression in case of ATCs, with almost doubled expression values in ATCs compared to normal samples (Table 1).

TABLE I. DEGs ANALYSIS SHOWS MAP3K8 EXPRESSION VALUES ACROSS ALL TYPES OF THYROID CANCER.

| Gene Symbol | Normal | ATC | PDTC |
|---|---|---|---|
| MAP3K8 | 2.745913 | 4.481513 | 3.450524 |

TABLE II. DEGs ANALYSIS SHOWS HRAS AND NRAS EXPRESSION VALUES ACROSS ALL TYPES OF THYROID CANCER.

| Gene Symbol | Normal | ATC | PDTC |
|---|---|---|---|
| HRAS | 7,318032 | 9,757364 | 8,545390 |
| NRAS | 7,245183 | 8,390965 | 6,616174 |

Even if this over-expression of H- and N-RAS is not as significant as the one of MAP3K8, it could be interesting to analyse any correlation between RAS and MAP3K8 up-regulation, since they have been associated with higher malignancy and shorter overall survival rate in PDTC [13], and their mutations are among the most common genetic alterations leading to tumors formation, along with BRAF mutation.

Stemness index evaluation showed a correlation between higher index values and MAP3K8 higher expression levels but this aspect needs to be further investigated in future.

MAP3K8 up-regulation in ATC samples could suggest possible correlations with some clinico-pathological parameters which are typical features of this tumor type. For example, as ATC is a rare and aggressive form of thyroid cancer that affects mainly elderlies displaying high prevalence of BRAF-V600E mutation, it is clear that a strong correlation between age and BRAF mutation exists. The possibility to detect this MAP3K8 over expression in previous tumor stages or in earlier ages could be a very powerful strategy for early diagnosis and prevention. Recent studies demonstrated that a sub-type of tumor cells showing stem cell like properties may harbor particularly aggressive phenotypes such as a strong therapy resistance and a high ability to metastasize. Since cancer progression involves the acquisition of stem-cell-like features we tried to evaluate the stemness degree of this particular thyroid cancer type: a high stemness degree corresponding to a massive presence of stem cell like features could be a predictive parameter of tumor outcome, and could give useful indications about its course Pathway analysis conducted on our DEGs list, showed that MAP3K8 is involved in MAPK signaling pathway, Toll-like receptor signaling pathway and T cell receptor signaling pathway.

Moreover, it is worth to mention how, among these pathways, a fundamental role of immune system dynamics is interestingly involved. Respectively, the activation of MAPK pathway, is crucial for transcriptional and nontranscriptional responses of the immune system [14]; toll-like receptor signaling pathway leads to activation of the transcription factors NF-κB and IRFs, which dictate the outcome of innate immune responses [15] and T cell receptor signaling pathway T-cell-receptor (TCR) signaling in response to antigen recognition has a central role in the adaptive immune response [16].

This demonstrates that MAP3K8 plays a decisive role in cell proliferation and in all the inflammatory events related to the tumor formation process and that a link between cancer and immune system is present.

## IV. CONCLUSIONS

The involvement of MAP3K8 in the formation and progression of thyroid cancer has been discussed in literature, as well as its contribution to vemurafenib resistance of mutant BRAF V600E thyroid cancer. Here we demonstrated that high values of MAP3K8 expression are related to a particularly aggressive and lethal tumor type, ATC. The possibility of considering MAP3K8 as a prognostic biomarker would be a great chance both to predict response to therapy and the progression of cancer itself but needs to be confirmed by future in vivo experiments. Moreover, functional variation of TLR4 genes is involved in chronic obstructive pulmonary disease and pulmonary tuberculosis: this makes MAP3K8 a possible target for investigating its potential roles both for diagnostic and prognostic biomarker in tuberculosis.

ACKNOWLEDGMENT

Authors of this paper acknowledge support from the STriTuVaD project. The STriTuVaD project has been funded by the European Commission, under the contract H2020-SC1-2017-CNECT-2, No. 777123. The information and views set out in this article are those of the authors and do not necessarily reflect the official opinion of the European Commission. Neither the European Commission institutions and bodies nor any person acting on their behalf may be held responsible for the use which may be made of the information contained therein.